%% file: priced_arxiv.tex
\documentclass{ifacconf}

\usepackage{graphicx}          
\usepackage{natbib}        
\usepackage{amsmath}
\usepackage{amssymb}
\usepackage{bm}
\usepackage{mathtools}
\usepackage{color,soul}
\usepackage{nicefrac}
\usepackage{hyphenat}
\usepackage{isomath}
\usepackage{algpseudocode}
\usepackage{algorithm}
\usepackage{outlines}
\usepackage{setspace}

\usepackage{tikz}
\input{tikzsettings.tex}

\newtheorem{model}[thm]{Model}

\newcommand*{\tran}{^{\mkern-1.5mu\mathsf{T}}\!}  

\DeclareMathOperator{\Tr}{Tr}

\def\d{\ensuremath{\mathrm{d}}}

\def\norm[#1]{\left|#1\right|}
\def\shortnorm[#1]{|#1|}

\def\Xs{\mathcal{X}}
\def\Ys{\mathcal{Y}}
\def\Us{\mathcal{U}}

\def\No{\mathbb{N}_{0}}
\def\N{\mathbb{N}}
\def\R{\mathbb{R}}
\def\S{\mathbb{S}}

\def\Ns{\mathcal{N}}

\def\Rs{\mathcal{R}}

\def\Ss{\mathcal{S}}
\def\Ks{\mathcal{K}}
\def\Qs{\mathcal{Q}}

\def\As{\mathcal{A}}  
\def\Ls{\mathcal{L}}  
\def\Lsc{\Ls_{\mathrm{c}}}  
\def\Es{\mathcal{E}}  
\def\Cs{\mathcal{C}}  
\def\Bs{\mathcal{B}}  
\def\Us{\mathcal{U}}  
\def\Usc{\Us_{\mathrm{c}}}  
\def\Usu{\Us_{\mathrm{u}}}  
\def\Esc{\Es_{\mathrm{c}}}  
\def\Esu{\Es_{\mathrm{u}}}  
\def\UscN{{\Usc}_\mathrm{N}}
\def\CsN{{\Cs}_\mathrm{N}}
\def\EsN{{\Es}_\mathrm{N}}
\def\IN{I_\mathrm{N}}
\def\cN{c_\mathrm{N}}

\def\nup{{n_{\mathrm{u}}}}
\def\nx{{n_{\mathrm{x}}}}

\def\xv{\boldsymbol{x}}

\def\xiv{\boldsymbol{\xi}}

\def\Am{\boldsymbol{A}}
\def\Bm{\boldsymbol{B}}

\def\I{\mathbf{I}}
\def\O{\mathbf{0}}

\def\Mm{\boldsymbol{M}}
\def\Nm{\boldsymbol{N}}

\def\Pm{\boldsymbol{P}}
\def\Qm{\boldsymbol{Q}}

\def\Rm{\boldsymbol{R}}

\def\Xm{\boldsymbol{X}}

\def\Km{\boldsymbol{K}}

\def\Idle{\mathtt{Idle}}
\def\InUse{\mathtt{InUse}}
\def\Bad{\mathtt{Bad}}
\def\true{\mathtt{true}}

\def\e{\mathrm{e}}

\def\edge#1{\xrightarrow{#1}}

\begin{document}

\begin{frontmatter}

\title{Scalable Traffic Models for Scheduling of Linear Periodic Event-Triggered Controllers}

\author[First]{Gabriel de A. Gleizer} 
\author[First]{Manuel Mazo Jr.} 

\address[First]{Delft Center for Systems and Control, TU Delft%
	, The Netherlands \\ (e-mail: \{g.gleizer, m.mazo\}@tudelft.nl)}

\thanks{This work is supported by the European Research Council through the SENTIENT project (ERC-2017-STG \#755953).}

\begin{abstract}

This paper addresses the problem of modeling and scheduling the transmissions generated by multiple event-triggered control (ETC) loops sharing a network. 
We present a method to build a symbolic traffic model of periodic ETC (PETC), which by construction provides an exact simulation of such traffic. The model is made in such a way as to avoid the combinatorial explosion that is typical of symbolic models in many applications. It is augmented with early triggering actions that can be used by a scheduler to mitigate communication conflicts. The complete networked control system is then modeled as a network of timed game automata, for which existing tools can generate a strategy that avoids communication conflicts, while keeping early triggers to a minimum.
By construction, our proposed symbolic model is a quotient model of the PETC. It is relatively fast to build, and it generates few to no spurious transitions. We finally demonstrate modeling and scheduling for a numerical example.
\end{abstract}


\end{frontmatter}

\section{INTRODUCTION}
Networks have become prevalent as the communication media for control devices. Despite the cost and implementability benefits brought by such Networked Control Systems (NCSs), the lack of dedicated communication lines for each sensor and actuator has introduced a challenge for practitioners: managing the communication required by each controller without compromising the control performance itself. In this context, aperiodic sampling methods such as Event-Triggered Control \cite[ETC,][]{tabuada2007event} and Self-Triggered Control \citep[STC,]{anta2008self} have been proposed. These methods significantly decrease network usage when compared to standard periodic sampling. ETC communications are triggered by events that are generally dependent on the system states. STC communication times are determined by the controller after every new data acquisition, generally by predicting when an ETC would trigger.\footnote{For an introduction on ETC and STC, see \cite{heemels2012introduction}.} Since then, many studies have focused on designing sampling strategies to reduce communication even further \citep[see, e.g.,][]{wang2008event, girard2015dynamic, dolk2017output}, among which there is periodic event-triggered control \citep[PETC,][]{heemels2013periodic}, which provides more practical implementations. Other researchers have proposed co-designing the controller and triggering mechanism to achieve the desired control performance \citep[e.g.,][]{peng2013event}, in some cases explicitly aiming at minimal communication \citep{donkers2014minimum}. We do not consider this co-design case in this work because we want to keep the concerns of control design separated from the digital implementation.

Despite the high communication savings achieved by ETC and STC, little research has addressed the coordination of data transfers from multiple controllers in a single network. Communication conflicts in an NCS can create stability issues, and scheduling is particularly difficult for ETC, since its triggering times vary immensely. Among the few previous research that addressed scheduling for ETC are \cite{kolarijani2016formal, mazo2018abstracted, fu2018traffic}, who propose a method to design conflict-free schedulers for ETC by means of symbolic abstractions of the ETC traffic. Using timed game automata (TGA) for approximately simulating ETC traffic, they demonstrate that a scheduling strategy can be computed by composing multiple traffic TGAs with a network TGA and solving a safety game. The major drawback of the abstractions presented in \cite{kolarijani2016formal} is the curse of dimensionality: their proposed isotropic partitioning creates a model with the number of locations that depend exponentially on the state-space dimension of the plant. For PETC, a traffic model was also proposed in \cite{fu2018traffic}, but it uses essentially the same method as \cite{kolarijani2016formal} and, thus, suffers from the same dimensionality issue.

In this paper, we follow the same philosophy of \cite{mazo2018abstracted} for scheduling, but propose a different way of creating the traffic models: instead of partitioning space, we partition time, and determine the states associated with a given triggering time \emph{a posteriori}. For PETC this allows to construct a quotient model \citep{tabuada2009verification}, which provides an exact simulation relation with the actual traffic generated. 
Generating the state space of the quotient model is relatively straightforward; the resulting regions are intersections of quadratic non-convex cones that, despite being easy to check membership online, pose a difficulty when determining the transition relations: it becomes a non-convex quadratic constrain satisfaction problem, which is in general NP-hard \citep{park2017general}. We propose using semidefinite relaxations \citep{boyd2004convex, park2017general}, which are fast and reliable, and seem to be remarkably tight for the problems we solve.
After having constructed the traffic model, we augment it to allow for controllable early triggers. These constitute the actions the scheduler has to avoid conflicts. In most ETC formulations, triggering earlier is safe in terms of stability, by the construction of the triggering mechanism. Finally, we follow the steps in \cite{mazo2018abstracted} to compose the scheduling problem, with some minor modifications to keep the number and earliness of scheduling interventions to a minimum level. For testing it, we generate strategies using UPPAAL Tiga \citep{behrmann2007uppaal} and provide simulation results for an NCS with two ETC loops. This demonstrates the usage of our method, which can support implementation of PETC in real NCSs, while helping realize the full potential of event-triggered control.

\subsection{Notation}

We denote $\No$ the set of natural numbers including zero, $\N \coloneqq \No \setminus \{0\}$, and $\R_+$ the set of non-negative reals. For a square matrix $\Am \in \R^{n \times n},$ we write $\Tr(\Am)$ to denote its trace, and $\Am \succ \O$ ($\Am \succeq \O$) if $\Am$ is positive definite (semi-definite). The sets $\S, \S_+$ and $\S_{++}$ are the sets of symmetric, positive definite, and positive semi-definite matrices, respectively. For a set $\Xs$, we denote by $\bar{\Xs}$ its complement; when $\Rs \subseteq \Xs \times \Xs$ is an equivalence relation on $\Xs$, we denote by $[x]$ an equivalence class of $x \in \Xs$, by $\Xs/\Rs$ the set of all equivalent classes, and by $\pi_{\Rs}(x): \Xs \to \Xs/\Rs$ the natural projection map taking a point $x \in \Xs$ to its equivalence class, that is, $\pi_{\Rs}(x) = [x] \in \Xs/\Rs$.

\section{PRELIMINARIES}

\subsection{Transition systems}

In order to be able to formally establish a relation between systems, we must introduce an overarching definition of what is a system, and how can a (finite-state) system model another (infinite-state) one. For that, we use the framework of \cite{tabuada2009verification}:

\begin{defn}[Transition System \citep{tabuada2009verification}]\label{def:system} 
A system $\Ss$ is a tuple $(\Xs,\Xs_0,\Us,\Es,\Ys,H)$ where:
	\begin{itemize}
		\item $\Xs$ is the set of states,
		\item $\Xs_0 \subseteq \Xs$ is the set of initial states,
		\item $\Us$ is the set of inputs,
		\item $\Es \subseteq \Ls \times \Us \times \Ls$ is the set of edges (or transitions),
		\item $\Ys$ is the set of outputs, and
		\item $H: \Xs \to \Ys$ is the output map.
	\end{itemize}
\end{defn}

A system is called finite (infinite) state if the cardinality of $\Xs$ is finite (infinite). A system is called autonomous if $\Us = \emptyset$, in which case  a transition is denoted by a pair $(x, x') \in \Xs\times\Xs$ instead of a triplet.

We aim at constructing an Automaton model of the timing of an ETC by using the notion of simulation relation:

\begin{defn}[Simulation Relation \citep{tabuada2009verification}]
	~\\Consider two systems $\Ss_a$ and $\Ss_b$ with $\Ys_a$ = $\Ys_b$. A relation $\Rs \subseteq \Xs_a \times \Xs_b$ is a simulation relation from $\Ss_a$ to $\Ss_b$ if the following conditions are satisfied:
	\begin{itemize}
		\item for every $x_{a0} \in \Xs_{a0}$, there exists $x_{b0} \in \Xs_{b0}$ with $(x_{a0}, x_{b0}) \in \Rs;$
		\item for every $(x_a, x_b) \in \Rs, H_a(x_a) = H_b(x_b);$
		\item for every $(x_a, x_b) \in \Rs,$ we have that $(x_a, u_a, x_a') \in \Es_a$ implies the existence of $(x_b, u_b, x_b') \in \Es_b$ satisfying $(x_a', x_b') \in \Rs.$
	\end{itemize}
\end{defn}

Whenever there is a simulation relation from $\Ss_a$ to $\Ss_b$, we use the notation $\Ss_a \subseteq \Ss_b$. Essentially, a simulation relation $\Rs \subseteq \Xs_a \times \Xs_b$ captures which states of $\Ss_a$ are simulated by which states of $\Ss_b$: for the right state selection, their outputs are the same; and every transition in $\Ss_a$ leads to a state whose output can also be attained in $\Ss_b$ after a single transition. It is important to notice, however, that there might be transitions in $\Ss_b$ that lead to states that are not related to the ones attained in $\Ss_a$. When using simulation relations to model the behavior of a system, these transitions are called spurious transitions.

Finally, we introduce the notion of quotient system, which is core to building symbolic models:

\begin{defn}[Quotient System \citep{tabuada2009verification}] Con-\\sider a system $\Ss = (\Xs,\Xs_0,\Us,\Es,\Ys,H)$ and let $\Rs$ be an equivalence relation on $\Xs$ such that $(x, x') \in \Rs \implies H(x) = H(x')$. The quotient of $\Ss$ by $\Rs$, denoted by $\Ss_{/\Rs},$ is the system $(\Xs_{/\Rs},$ $\Xs_{/\Rs0},$ $\Us,$ $\Es_{/\Rs},$ $\Ys,$ $H_{/\Rs})$ consisting of
	\begin{itemize}
		\item $\Xs_{/\Rs} = \Xs{/\Rs}$;
		\item $\Xs_{/\Rs0} = \{x_{/\Rs} \in \Xs_{/\Rs}: x_{/\Rs} \cap \Xs_0 \neq \emptyset\}$;
		\item $(x_{/\Rs}, u, x'_{/\Rs}) \in \Es_{/\Rs}$ if there exists $(x,u,x') \in \Es$ with $x \in x_{/\Rs}$ and $x' \in x'_{/\Rs}$;
		\item $H_{/\Rs}(x_{/\Rs}) = H(x)$ for some $x \in x_{/\Rs}.$
	\end{itemize}
\end{defn}

Building a quotient system is fundamentally aggregating states of the original system that produce the same output, and then determining the transitions so that every possible transition of the original system is reproduced in the quotient (symbolic) system. By construction, $\Ss \subseteq \Ss_{/\Rs}$.

\subsection{Timed automata}

Timed Automata are regular Automata that make use of clocks, which are resettable real-valued variables measuring the passage of time. Let $\Cs$ be a finite set of said clocks, and consider $\bowtie \in\! \{{<,} {\leq,} {=,} {\geq,} >\}$. A clock constraint $g$ is a conjunctive formula of atomic constraints $c \bowtie k, c \in \Cs, k \in \N$. We denote by $\Bs(\Cs)$ the set of all clock constraints.
\begin{defn} (Timed Safety Automaton, \citep{bengtsson2003timed}). A Timed Safety Automaton is a tuple $\As = (\Ls,\Ls_0,\Us,\Cs,\Es,I)$ where:
	\begin{itemize}
		\item $\Ls$ is the finite set of locations (or discrete states),
		\item $\Ls_0 \subseteq \Ls$ is the set of initial locations,
		\item $\Us$ is the finite set of actions,
		\item $\Cs$ is the finite set of clocks,
		\item $\Es \subseteq \Ls \times \Bs(\Cs) \times \Us \times 2^\Cs \times \Ls$ is the set of edges (or transitions), and
		\item $I: \Ls \to \Bs(\Cs)$ assigns invariants to locations.
	\end{itemize}
\end{defn}
A TSA is a system with both discrete states (the locations) and continuous states (the clocks). All clocks increase value at the same rate, but some transitions can reset the value of certain clocks. The system can change locations through edges, depending on the action taken and the clock's values.
We denote by $l \edge{g,a,r} l'$ the transition from $l\in\Ls$ to $l'\in\Ls$ under action $a\in\Us$, with $r \subseteq \Cs$ as the set of clocks reset when this transition is taken, and $g$ over $\Cs$ as the guards that enabled the transition. \emph{Invariants} of a location are the sufficient clock conditions for a transition to happen; in other words, the system is forced to leave the place $l$ if a clock $c$ violates any invariant $I(l)$. On the other hand, a \emph{guard} of an edge is a necessary condition for it to take place.

TGA extend TSA by partitioning the set of actions into controllable and uncontrollable. Controllable actions are decisions that the system operator can choose, while uncontrollable actions are taken independently of the system operator (e.g., by the environment or an opponent).
\begin{defn} (Timed Game Automaton, \citep{bengtsson2003timed}). A Timed Game Automaton is a tuple $\As = (\Ls,\Ls_0,\Usc,\Usu,\Cs,\Es,I)$ where:
	\begin{itemize}
		\item $(\Ls,\Ls_0,\Usc \cup \Usu,\Cs,\Es,I)$ is a TSA,
		\item $\Usc$ is the set of controllable actions,
		\item $\Usu$ is the set of uncontrollable actions, and
		\item $\Usc \cap \Usu = \emptyset$.
	\end{itemize}
\end{defn}
The distinction between controllable and uncontrollable is paramount in our case. The scheduler can control when to sample, but not how the system will react to this choice.

It is important to introduce the notion of a game strategy. Let $\As$ be a TGA, and $\Lsc\subseteq\Ls$ be its set of locations, for which a controllable action exists. A strategy $S:\Lsc\times\Cs \to 2^{\Usc}$ determines which actions can be taken depending on the location the automaton is and on its clocks' values. A deterministic strategy outputs a single action, while a time-invariant strategy takes only locations as inputs.

Finally, for scheduling, we will need to combine models of multiple control systems among themselves and with a model of the network. TGAs can be combined into a \emph{network of timed game automata} (NTGA), which allows for modularity \citep{bengtsson2003timed}. An NTGA consists of $n$ timed game automata $\As_i = (\Ls_i,\Ls_{i0},\Usc,\Usu,\Cs,\Es_i,I_i)$ where the set of actions over the network is defined in such a way that uncontrollable actions take precedence over controllable actions. Additionally, a location of the network, denoted as $\bar{l} \coloneqq (l_1, ..., l_n)$, has its invariant $I(\bar{l}) = \wedge_iI_i(l_i)$. Most importantly, TGAs within an NTGA can have transitions influence each other through \emph{synchronization channels}: for a channel {\tt a}, the initiating transition is labeled {\tt a!} and, when fired, all transitions labeled {\tt a?} have to fire simultaneously.

\subsection{Periodic event-triggered control}

Consider the plant with a sample-and-hold state-feedback control below:
\begin{align}
\dot{\xiv}(t) &= \Am\xiv(t) + \Bm\Km\hat{\xiv}(t),\label{eq:plant}\\
\xiv(0) &= \hat{\xiv}(0) = \xiv_0, \nonumber
\end{align}
where $\xiv(t) \in \R^\nx$ is the state with initial value $\xiv_0$, $\hat{\xiv}(t) \in \R^\nx$ is the available measurement of the state, $\Km\hat{\xiv}(t) \in \R^\nup$ is the control input, and $\Am, \Bm, \Km$ are matrices of appropriate dimensions. Assume $\Km$ is a matrix designed so that the system is asymptotically stable when $\hat{\xiv} \equiv \xiv$. 
The controller above uses a zero-order hold mechanism; i.e., consider a sequence of sampling times $t_i \in \R_+$, with $t_0 = 0$ and $t_{i+1} - t_i > \varepsilon$ for some $\varepsilon > 0$. Then $\hat{\xiv}(t) = \xiv(t_i), \forall t \in [t_i, t_{i+1})$.

In event-triggered control, the sequence of times $t_i$ is generated by a \emph{triggering condition}, which is generally a function of the states of the system. In periodic ETC, such a condition is checked periodically, with a fundamental checking period $h$:
\begin{equation}\label{eq:quadtrig}
t_{i+1} = \inf\left\{t=kh>t_i, k \in \N \middle|
\!\begin{array}{c}
\begin{bmatrix}\xiv(t) \\ \xv\end{bmatrix}\tran
\Qm \begin{bmatrix}\xiv(t) \\ \xv\end{bmatrix} > 0 \\
\vee \ t-t_i \leq \bar{k}h\phantom{\dot{\hat{I}}}
\end{array}\!
\right\},
\end{equation}
where $\xv = \xiv(t_i)$, $\Qm \in \S^{2\nx}$ is the designed triggering matrix, and $\bar{k}$ is a chosen maximum inter-event time. Many of the triggering conditions available in the literature can be written as in Eq.~\eqref{eq:quadtrig}. We kindly refer the interested reader to \cite{heemels2013periodic} for the list of conditions and their formulations.

In-between $t_i$ and $t_{i+1}$, the value of $\xiv(kh)$ can be precisely determined as
\begin{equation}\label{eq:Mk}
\xiv_{\xv}(kh) = \Mm(k)\xv, \quad \Mm(k) \coloneqq \e^{\Am kh} + \int_0^{kh}\e^{\Am\tau}\d\tau \Bm\Km,
\end{equation}
where $\xiv_{\xv}(t)$ is used to denote the value of $\xiv$ at $t$ when $\xiv(0) = \hat{\xiv}(t) = \xv$. One can determine the discrete inter-event $\kappa \coloneqq (t_{i+1}-t_i)/h$ time as a function of the currently held state by combining Equations \eqref{eq:quadtrig} and \eqref{eq:Mk}:
\begin{equation}\label{eq:petc_time}
\begin{gathered}
\kappa(\xv) = \min\left\{k \in \{1, 2, ...\bar{k}\}\middle| \xv\tran\Nm(k)\xv > 0 \vee k=\bar{k}\right\} \\
\Nm(k) \coloneqq \begin{bmatrix}\Mm(k) \\ \I\end{bmatrix}\tran
\Qm \begin{bmatrix}\Mm(k) \\ \I\end{bmatrix},
\end{gathered}
\end{equation}
where $\I$ denotes the identity matrix.

\section{PROBLEM FORMULATION}

The starting point for scheduling ETC traffic is predicting the timing of its communications. For doing so, one can try to construct a model of such timing. Inspired by \cite{kolarijani2015traffic, mazo2018abstracted}, we use symbolic abstractions; however, we aim to build a quotient model, such that an exact simulation relation is obtained. More than that, we want to mitigate the curse of dimensionality that is typical of such abstractions:

\begin{prob}\label{prob:theproblem}
Build a quotient model $\Ss_{/\Rs}$ for the traffic generated by system \eqref{eq:plant} using triggering condition \eqref{eq:quadtrig} in such a way that the cardinality of $\Xs_{/\Rs}$ does not depend directly on $\nx$.
\end{prob}

A traffic model alone is not sufficient for scheduling purposes. System \eqref{eq:plant} is autonomous,  and a scheduler needs to be able to alter the traffic pattern in some way in order to avoid communication conflicts. We choose to allow the scheduler to request data before the ETC triggers. Therefore, we need to enrich the traffic model with controllable actions that represent this early triggering:

\begin{prob}\label{prob:model}
	Enhance $\Ss_{/\Rs}$ with transitions that capture the evolution of system \eqref{eq:plant} when inter-event times smaller than $\kappa(x)$ are chosen.
\end{prob}

Finally, we need to pose the scheduling problem. With the model of multiple event-triggered loops, as well as a model of the network, we can build a network of timed game automata that represents the complete NCS:

\begin{prob}\label{prob:solution}
	Design an NTGA that forms the scheduling problem, for which a strategy serves as a scheduler for the NCS with multiple event-triggered loops. In doing so, try to keep the number of communications to a small level.
\end{prob}

\section{PETC TRAFFIC MODEL}

Constructing a quotient model of the traffic generated by \eqref{eq:plant}--\eqref{eq:quadtrig} requires two steps: 1) gathering the states that share the same output in a single quotient state, and 2) computing the transition relations between such quotient states. Before that, let us use Definition \ref{def:system} to define the precise, infinite-state traffic model: it is the system $\Ss = (\Xs, \Xs_0, \emptyset, \Es, \Ys, H)$ where
\begin{equation}\label{eq:original}
\begin{aligned}
	\Xs &= \Xs_0 = \R^{\nx}; \\
	\Es &= \{(\xv, \xv') \in \Xs \times \Xs| \xv' = \xiv_{\xv}(h\kappa(\xv))\}; \\
	\Ys &= \{1, 2, ..., \bar{k}\}; \\
	H &= \kappa.
\end{aligned}
\end{equation}

We can now proceed to building the quotient model.

\subsection{Quotient state set}

Gathering states that share the same output is in a sense straightforward in PETC. From Eq.~\eqref{eq:petc_time}, we can determine the set $\Ks_k \subseteq \R^{\nx}$ of states that will certainly have triggered by time $k$:
\begin{equation}\label{setk}
\Ks_k = \begin{cases}
\{\xv \in \R^{\nx}| \xv\tran\Nm(k)\xv > 0\}, & k < \bar{k}, \\
\R^{\nx}, & k = \bar{k}.
\end{cases}
\end{equation}
To determine the state set whose output $k$ is the \emph{minimum} that satisfies $\xv\tran\Nm(k)\xv > 0$, all one needs to do is remove from $\Ks_k$ all states that could have triggered before, i.e., that belong to some $\Ks_{j}$ with $j < k$. This is expressed with the state $\Qs_k$ computed recursively as
\begin{equation*}
\Qs_k = \begin{cases}
\Ks_k \setminus \bigcup_{j=1}^{k-1}\Qs_j, & k > 1, \\
\Ks_k, & k = 1,
\end{cases}
\end{equation*}
which can be expressed as
\begin{equation}\label{eq:setq}
\Qs_k = \Ks_k \cap \bigcap_{j=1}^{k-1} \bar{\Ks}_{j}.
\end{equation}

By construction, $\Qs_k, k \in \{1, 2, ..., \bar{k}\}$ constitutes a partition of $\R^{\nx}$; also, $H(\xv) = k, \forall \xv \in \Qs_k$. Therefore, $\Xs_{/\Rs} = \{\Qs_1, \Qs_2, ...\}$ is a good candidate for a quotient state set of the system $\Ss$. Finally, different from \cite{kolarijani2016formal}, we have that $|\Xs_{/\Rs}| = \bar{k}$, i.e., the cardinality of the quotient state space does not depend explicitly on $\nx$. This in part accomplishes solving Problem \ref{prob:theproblem}; however, for completing the model, we need to establish the transitions between these quotient states. 
 
\begin{rem}
Matrices $\Nm(k)$ can be computed offline. Online determination of which region the current state $\xv$ belongs to requires at most $\bar{k}$ quadratic operations.
\end{rem}

\begin{rem}\label{rem:miet} Unperturbed state-feedback ETC has an intrinsic positive minimum inter-event time (MIET), which, in the case of PETC, can be bigger than $k=1$. In this case, for all $k < \underline{k}$, where $\underline{k}$ is such MIET, all $\Nm(k) \preceq \O.$ This can be checked offline, and the corresponding matrices may be discarded. Likewise, a maximum inter-event time $\bar{k}$ can naturally show up if, for some $k^*$, $\Nm(k^*) \succ \O$, which can also be checked offline. In this case, take $\bar{k} = k^*.$
\end{rem}

\subsection{Quotient transition relations}

The problem of determining the transition relation between two quotient states $\Qs_i$ and $\Qs_j$ is, from Eq.~\eqref{eq:original},
\begin{equation}\label{eq:transproblem}
\exists \xv \in \R^{\nx}: \xv \in \Qs_i, \xiv_{\xv}(ih) = \Mm(i)\xv \in \Qs_j,
\end{equation}
where the last equality uses Eq.~\eqref{eq:Mk}. Expanding $\Qs_i, \Qs_j$ with Eqs.~\eqref{eq:setq} and \eqref{setk} arrives in the following \emph{non-convex quadratic constraint satisfaction problem}:
\begin{equation}\label{eq:transqcqp}
\begin{aligned}
\exists \ \ & \xv \in \R^{\nx} \\
\text{s.t.} \ \ & \xv\tran\Nm(i)\xv > 0, \\
                 & \xv\tran\Nm(i')\xv \leq 0, \forall i' \in \{1,...,i-1\}, \\
                 & \xv\tran\Mm(i)\tran\Nm(j)\Mm(i)\xv > 0, \\
                 & \xv\tran\Mm(i)\tran\Nm(j')\Mm(i)\xv \leq 0, \forall j' \in \{1,...,j-1\}.
\end{aligned}
\end{equation}
The non-convexity of this problem can be easily checked using the facts that both $>$ and $\leq$ inequalities are present, and that the matrices $\Nm(i)$ are non-definite.\footnote{See Remark \ref{rem:miet}: the definite cases are discarded.} This can, on a first sight, be regarded as a disadvantage with respect to the model proposed in \cite{kolarijani2016formal}, whose (power) quotient states are convex; however, convex relaxations such as the semi-definite relaxation (SDR) from \cite{boyd2004convex} can be used. Additionally, we relax the strict inequalities with non-strict ones, so that it can fit the semi-definite programing formulation. The SDR becomes
\begin{equation}\label{eq:transsdr}
\begin{aligned}
\exists \ \ & \Xm \in \S_{+}^{\nx} \\
\text{s.t.} \ \ & \Tr(\Xm\tran\Nm(i)) \geq 0, \\
& \Tr(\Xm\Nm(i')) \leq 0, \forall i' \in \{1,...,i-1\}, \\
& \Tr(\Xm\Mm(i)\tran\Nm(j)\Mm(i)) \geq 0, \\
& \Tr(\Xm\Mm(i)\tran\Nm(j')\Mm(i)) \leq 0, \forall j' \in \{1,...,j-1\}, \\
& \Tr(\Xm) = 1,
\end{aligned}
\end{equation}
where the last equation was added to avoid the trivial solution $\Xm=0$; the value $1$ was chosen arbitrarily, since the problem without this constraint is homogeneous. User-friendly interfaces, such as CVX \citep{cvx}, and efficient solvers for semidefinite programming, such as SCS \citep{ocpb:16, scs} can be used for this problem. To determine (offline) the complete transition set $\Es_{/\Rs}$, one requires solving $\bar{k}^2$ semidefinite problems. The final model follows:
\begin{model}[PETC Traffic Model]\label{model0} The model is the system $\Ss_{/\Rs} = (\Xs_{/\Rs}, \Xs_{/\Rs0}, \emptyset, \Es_{/\Rs}, \Ys, H_{/\Rs})$ with
	\begin{itemize}
		\item $\Xs_{/\Rs} = \Xs_{/\Rs0} = \{\Qs_1, \Qs_2, ..., \Qs_{\bar{k}}\}$;
		\item $\Es_{/\Rs} = \{(\Qs_i, \Qs_j)| \text{Eq.~\eqref{eq:transsdr} is satisfied}\}$;
		\item $H_{/\Rs}(\Qs_k) = k.$
	\end{itemize}
\end{model}

By construction, we obtain the following result:
\begin{thm}\label{prop} Model \ref{model0} is a quotient system of $\Ss$ from Eq.~\eqref{eq:original}, and, therefore, $\Ss_{/\Rs}$ simulates $\Ss$. \end{thm}
In other words, all sequences of triggering times generated by system \eqref{eq:plant}--\eqref{eq:petc_time} can be generated by our model $\Ss_{/\Rs}$. This solves Problem \ref{prob:theproblem}.

\begin{rem} A relaxation generally provides conservative solutions. In our case, this means that we may find spurious transitions. If such transitions do occur, this does not change the fact that the constructed symbolic model simulates $\Ss$.
\end{rem}

\section{Scheduling of PETC systems}

\subsection{Early triggering and TGA}

As stated earlier, for the traffic model to be applicable for scheduling, we need to augment it with controllable transitions that correspond to early triggering. From a quotient state $\Qs_i$, one can allow early triggers for any $k\in\N: k < i$; for simplicity we choose to label the corresponding actions by $k$. It remains necessary to verify which transitions exist for such actions. Obviously, this can be done by solving the SDR problem \eqref{eq:transsdr} as before, replacing $j$ by $k$. We denote the set of early triggering transitions by $\Es^*$ and the resulting system as $\Ss_{/\Rs}^*$. Computing all of its transitions requires solving
$\bar{k} + 2\bar{k} + ... + \bar{k}(\bar{k}-1) = \bar{k}^2(\bar{k}-1)/2$ semidefinite problems.

Finally, we transform the quotient system into a TGA. For the game part, we set the early triggering actions in $\Ss_{/\Rs}^*$ as controllable, and the event triggers as controllable. All that is left is defining the clock set, the guards, and the invariants, resulting in the following TGA:
\begin{model}[PETC Traffic Timed Game]\label{model1} The model is the TGA $\As = (\Xs_{/\Rs}, \Xs_{/\Rs0}, \Usc, \Usu, \Cs, \Esc \cup \Esu, I)$ where
\begin{itemize}
	\item $\Usc = \{\mathtt{early}\}$;
	\item $\Usu = \{\mathtt{trigger}\}$;
	\item $\Cs = \{c\}$;
	\item $\Esc = \{(\Qs_i,c=k, \mathtt{early},\{c\}, \Qs_j): (\Qs_i,k,\Qs_j) \in \Es^*\}$; 
	\item $\Esu = \{(\Qs_i,c=i, \mathtt{trigger},\{c\}, \Qs_j): (\Qs_i,\Qs_j) \in \Es_{/\Rs}\}$;$\!$
	\item $I(\Qs_i) = (c \leq i).$
\end{itemize}
\end{model}

Model \ref{model1} requires some explanation related to clocks. First, we use one clock, that is reset at every transition. The invariant of a quotient state $\Qs_i$ is naturally $c\leq i$, because $i$ is the time that a trigger is sure to occur; hence $c=i$ is the clock constraint associated with this uncontrolled action. For the controlled, early triggering actions, the transition is enabled at discrete instants satisfying $c=k$, for $k<i$.

\subsection{Network and NCS models}

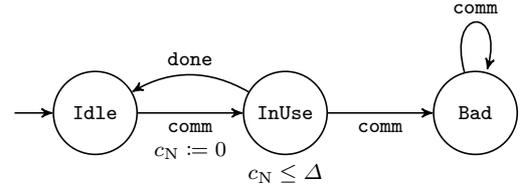
\begin{figure}
	\begin{center}
		\begin{tikzpicture}[->,>=stealth',auto,node distance=2.5cm,
		every node/.append style={font=\small}, semithick]
		\node[draw, circle, minimum size=1.1cm] 		(i)                    {$\Idle$};
		\node[]						(a) [left=0.5cm of i]			{};
		\node[draw, circle, minimum size=1.1cm]         (u) [right of=i] 	   {$\InUse$};
		\node[draw, circle, minimum size=1.1cm]         (b) [right of=u] 	   {$\Bad$};
		
		\node[below=0.0cm of u] {$\cN \leq \Delta$};
		
		\path 	(a) edge						 (i)
				(i)	edge				node[below,text width=1cm,align=center] {{\tt comm} \\ $\cN \coloneqq 0$}  (u)
				(u)	edge[bend right]	node[above] {\tt done} (i)
					edge				node[below] {\tt comm} (b)
				(b) edge[loop above]	node[above] {\tt comm} (b);
		\end{tikzpicture}
		\caption{\label{fig:net} {TGA of a shared network.}}
	\end{center}
\end{figure}

For scheduling, we follow the same strategy as described in \cite{mazo2018abstracted}. First, we need a model of the network. As in \cite{mazo2018abstracted}, the model must capture a channel occupancy time; while the network is being used, the scheduler must avoid that a second communication happens. We use the same model of network as they use, with a minor technical change:
\begin{model} (Network TGA, adapted from \cite{mazo2018abstracted}).\label{model:net} The model is the TGA $\Ns = (\Ls, l_0, \UscN, \emptyset, \CsN, \EsN,$ $\IN)$ where
	\begin{itemize}
		\item $\Ls = \{\mathtt{Idle}, \mathtt{InUse}, \mathtt{Bad}\}$;
		\item $\UscN = \{\mathtt{comm}, \mathtt{done}\}$;
		\item $\Cs = \{\cN\}$;
		\item $\EsN = \{{(\Idle,\true,\mathtt{comm},\{\cN\}, \InUse)},$ \\
			${(\InUse,\cN=\Delta,\mathtt{done},\emptyset, \Idle)}, $ \\
			$(\InUse,\true,\mathtt{comm},\emptyset, \Bad),$ \\
			$(\Bad,\true,\mathtt{comm},\emptyset, \Bad) \}$; 
		\item $\IN(\InUse) = (\cN \leq \Delta),$
	\end{itemize}
where $\Delta$ is the maximum channel occupancy time.
\end{model}
The difference of this model with respect to \cite{mazo2018abstracted} is that, here, all actions are controlled. We do this because of how NTGA are composed in UPPAAL Tiga: if an uncontrolled edge is synchronized with a controlled edge, the composed edge is uncontrolled. When we compose the traffic models with the network model, we want the {\tt early} communications to be controlled, and the {\tt trigger} ones not to. Model \ref{model:net} is represented in Fig.~\ref{fig:net}.
To model the NCS, we build an NTGA of the two or more traffic models $\As_i$ with the network model $\Ns$. What remains to be done is synchronizing the correct actions. For this, we add a synchronization channel called {\tt up}, which is used as follows:
\begin{itemize}
	\item every {\tt early} and {\tt trigger} actions of each traffic model $\As_i$ fires the synchronizing action {\tt up!};
	\item every {\tt comm} action of the network model $\Ns$ takes the synchronizing action {\tt up?}.
\end{itemize}
While avoiding the {\tt Bad} state is necessary, we also want that the number of early triggers is kept to a minimum, so as to benefit from the communication savings of ETC. For that, we introduce an integer variable $e, 0 \leq e \leq E$, representing an accumulated ``earliness'' of communications, with $E$ as the maximum allowed earliness.  It is essentially a bounded integrator that increases every time an early trigger is done and decreases when a natural trigger happens. It starts at zero and is updated as
\begin{equation}\label{eq:earliness}
e \gets \max(0, \min(E, e + r(k-i)-\bar{e}))
\end{equation}
for every {\tt trigger} or {\tt early} transition from any traffic model, from quotient state $\Qs_i$ when $c=k$. The parameters $r \in \N_+$ and $\bar{e}\in\N_+$ represent the cost of a time unit and a reference value for $e$, respectively. The earlier the trigger is, the higher the cost incurred. Parameter $\bar{e}$ is necessarily positive so as to allow that natural triggers discount $e$; otherwise, no safe solution is possible unless no early trigger ever occurs. Like any arithmetics on bounded integers, the evolution of $e$ can be represented as an automaton itself. UPPAAL Tiga allows one to use integer variables, and it performs the necessary operations automatically.

As a final note, it is important to remember that the model $\As$ uses a normalized time with respect to the check time $h$. It is not necessary to have the check times of every control loop to be the same, since TGA clocks take rational values; still, one needs to put the clocks and their constraints in the same time scale prior to composing the NCS model.

\subsection{Strategies for schedulers}

In UPPAAL Tiga, strategies can be generated so as to guarantee certain specifications. We refer the reader to the manual of UPPAAL Tiga \citep{behrmann2007uppaal} for the complete list. In our case, we want that the NTGA never enters state $\Bad$ of $\Ns$, while keeping the earliness below a certain threshold $E$. This can be achieved by setting the specification {\tt strategy safe = control: A[] not network.Bad and e < E}. The resulting strategy maps the locations of each automaton and their clock valuations into the decision of whether to trigger early or not. Therefore, a scheduler that implements such strategy needs to determine online the regions $\Qs_i$ that the state of each system belongs to, and keep track of how much time elapsed since the last communication of each plant.


\section{NUMERICAL RESULTS}

Consider two copies of a linearized batch reactor, taken from \cite{donkers2011thesis}, of the form \eqref{eq:plant} with
\begin{equation}\label{eq:AB}
\begin{aligned}
\Am_i &= \begin{bmatrix}1.38 & -0.208 & 6.715 & -5.676 \\
-0.581 & -4.29 & 0   &  0.675 \\
 1.067 &  4.273 & -6.654 &  5.893 \\
 0.048 &  4.273 &  1.343 & -2.104\end{bmatrix}, \\
\Bm_i &= \begin{bmatrix} 0  &  0  \\
5.679 & 0 \\
1.136 & -3.146 \\
1.136 &  0\end{bmatrix}, \quad i \in \{1, 2\}.
\end{aligned}
\end{equation}
Two different controllers $\Km_i$ were designed for this plant using LQR with matrices $\Qm_{\text{LQR},1} = \Qm_{\text{LQR},2} = \I$ and $\Rm_1 = 0.1\I, \Rm_2 = 0.05\I$. The Lyapunov function chosen was the LQ cost, that is, setting $\Qm_{\text{lyap},i} = \Qm_{\text{LQR},i} + \Km_i\tran\Rm_i\Km_i$ and solving the continuous-time Lyapunov equation for $\Pm_i$. We used a triggering condition based on the Lyapunov function, so as to guarantee that
$$ \dot{V}_i(t) \leq -\rho_i\xiv_i(t)\tran\Pm_i\xiv_i(t), $$
for some $0 < \rho_i < 1$.  We set $\rho_1 = \rho_2 = 0.8.$ This triggering condition can be expressed in quadratic form \eqref{eq:quadtrig}, after doing the necessary algebraic manipulations, with
$$ \Qm_i = \begin{bmatrix}\Am_i\tran\Pm_i + \Pm_i\Am_i + \rho_i\Qm_{\text{lyap},i} & \Pm_i\Bm_i\Km_i \\ \Km_i\tran\Bm_i\tran\Pm_i & \O\end{bmatrix}. $$
In both cases, we set $h_1=h_2=h=0.01$ and $\bar{k} = 20$; however, following Remark \ref{rem:miet}, we obtained natural maximum inter-event times at $\bar{k}_1 = 19$ and $\bar{k}_2 = 16$ by imposing that $\Nm(k)$ have its largest eigenvalue bigger than $10^{-3}$. Likewise, both have MIETs greater than 1: $\underline{k}_1 = 6$, $\underline{k}_2 = 4.$ 

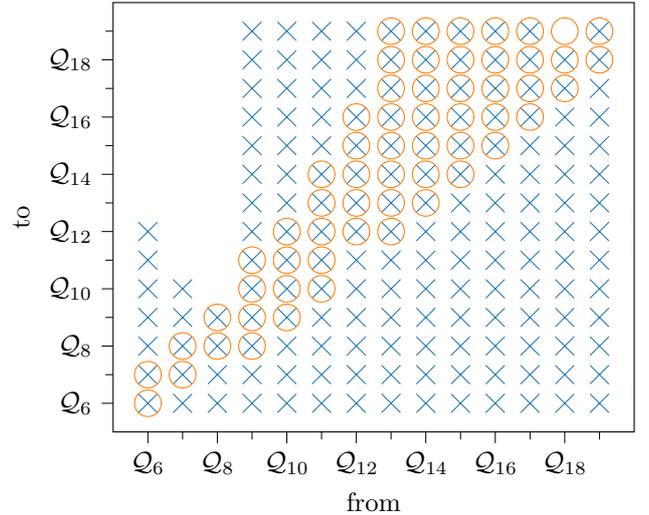
\begin{figure}
	\begin{center}
		\input{transition.tex}
		\caption{\label{fig:transition} {Transition relations of $\Ss_{/\Rs}^*$ of loop 1, for {\tt trigger} actions (x) and {\tt early} actions (o) with $k=1$.}}
	\end{center}
\end{figure}

To build Model \ref{model1} for each control loop, we used Python with Numpy, Scipy and control packages, and CVXPY \citep{cvxpy, cvxpy_rewriting} with solver SCS \citep{ocpb:16, scs} to solve the semidefinite problems involved. The whole process of computing matrices $\Nm(k)$ and solving the semidefinite problems took 46.64 seconds for loop 1 and 31.51 seconds for loop 2. The computer used is a MacBook Pro with a 3.1 GHz Intel Core i5 CPU and memory of 8 GB, 2133 MHz LPDDR3. The resulting transition relation for closed-loop system 1 is represented in Figure \ref{fig:transition}. As one can see, there is a significant amount of nondeterminism introduced by this model, especially for high triggering times.

A series of scripts was used to generate the XML files that are used for TGA models in UPPAAL Tiga. We used all times in the NTGA relative to $h$, and set $\Delta = 1$. The earliness parameters for Eq.~\ref{eq:earliness} were $r=2, \bar{e}=1, E=2.$ These parameters allow the scheduler to trigger one step earlier at every two communications.

The strategy was solved in UPPAAL STRATEGO \citep{david2015uppaal} version 4.1.20-5, which includes all functionalities of UPPAAL Tiga. It took 0.864 s to find a solution. The generated strategy is too long to be reproduced in this paper, but we give below one example of when an early trigger has to occur:
\begin{align*} \text{If } & \text{System 1 is in } \Qs_6 \text{, System 2 is in } \Qs_4 \text{, and } e=0, \\ 
&\text{when } 
c_1 = 5 \text{ and } c_2 \in \{1, 2, 3\}, \text{ do {\tt early} on System 1;} \\
&\text{when } 
c_2 = 3 \text{ and } c_1 \in \{3, 4, 5\}, \text{ do {\tt early} on System 2,} 
\end{align*}	
where $c_i$ represents the clock valuation of system $i$. As one can see, the strategy is not deterministic. In the example above, the early trigger can be executed on any of the loops when $(c_1,c_2) = (5,3)$. In such case, the scheduler must arbitrate who triggers.

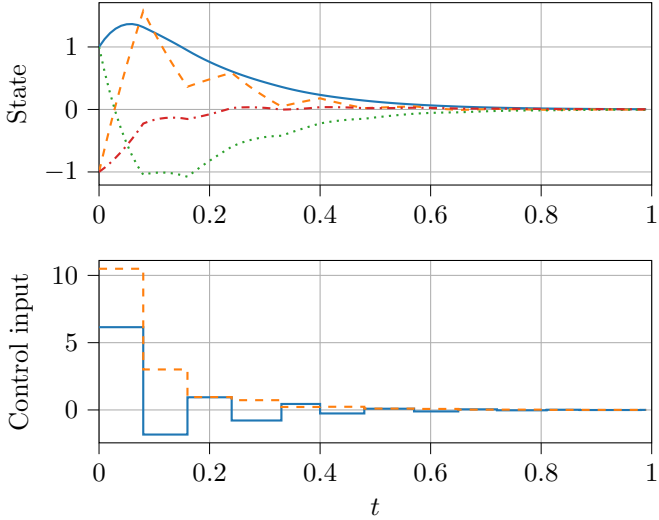
\begin{figure}
	\begin{center}
		\input{out1.tex}
		\caption{\label{fig:out1} {Trajectories of $\xiv_1(t)$ (top) and $\Km_1\hat{\xiv}_1(t)$ (bottom).}}
	\end{center}
\end{figure}

Figures \ref{fig:out1} and \ref{fig:out2} show the results of a simulation of the two control loops executing in parallel with the communication managed by the synthesized scheduler. The initial conditions are $\xiv_1(0) = \begin{bmatrix}1 & -1 & 1 & -1\end{bmatrix}\tran$ and $\xiv_2(0) = \begin{bmatrix}1 & 2 & 3 & 4\end{bmatrix}\tran$. The first pair of communications were arbitrated on a round-robin fashion. Figure \ref{fig:comm} shows the communication pattern of the NCS. As we can see, both systems' states converge to zero, while there is no conflict in communications. As designed through the earliness mechanism, about half of the communications are early triggers, and half are natural, event triggers.

\begin{figure}
	\begin{center}
		\input{out2.tex}
		\caption{\label{fig:out2} {Trajectories of $\xiv_2(t)$ (top) and $\Km_2\hat{\xiv}_2(t)$ (bottom).}}
	\end{center}
\end{figure}
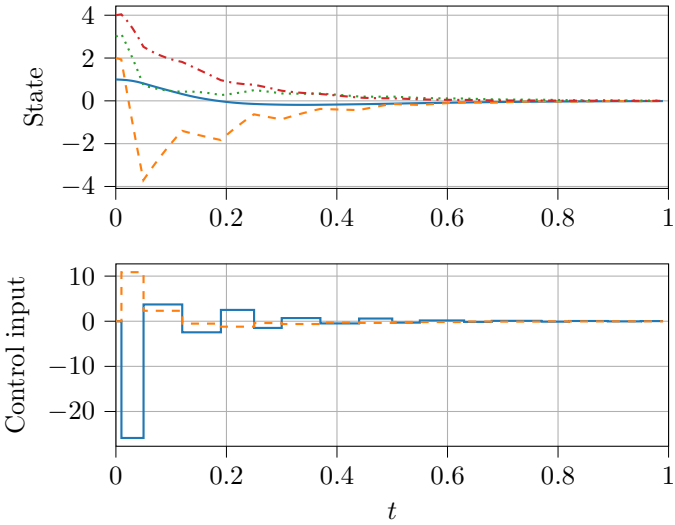
\begin{figure}
	\begin{center}
		\input{trigger.tex}
		\caption{\label{fig:comm} {Communication pattern of the simulated NCS: `x' marks represent event triggers, while `o' marks represent early triggers.}}
	\end{center}
\end{figure}
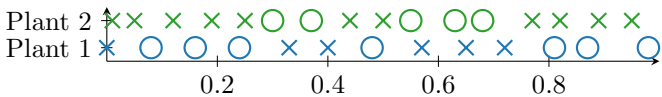
\section{CONCLUSIONS}

In this paper, we presented a method to build a quotient model of the traffic generated by PETC, and how to augment it and use it for scheduling of multiple PETC loops.
The quotient model has many advantages with respect to related work: first, it provides an exact simulation instead of an approximate one; and second, it avoids the combinatorial explosion created by isotropic partitioning of the state space. The state space and output map of the quotient model can be easily created straight from the PETC and system matrices, requiring no solution of LMIs or other optimization problems. The transition relations do require semidefinite problems to be solved, but only one per transition, with no reachability tools required. It is relatively fast to compute, and the models generated are reasonably small. The use of TGA models for scheduling of ETC had already been demonstrated in \cite{mazo2018abstracted}; here, we demonstrate that they can also be done for PETC, and argue that it is in fact simpler to do so.

Among the disadvantages of our solution is the high nondeterminism of the generated models. The state-space partitions are based solely on the output function, and each region seems to be large enough that, after some time, many regions can be reached. A highly nondeterministic traffic model can hamper the generation of strategies, as the predictability of the model after multiple steps gets smaller. One solution we are exploring is partitioning the regions further using backwards reachability.
A second disadvantage of this approach, shared with \cite{mazo2018abstracted}, is that the size of the NTGA state space grows exponentially with the number of control loops. This can make solving the scheduling problem impracticable. Solving strategies for TGA is EXPTIME-complete \citep{asarin1998controller}, so controlling the size of the (N)TGA is paramount. Methods to do so are subject of future research. 
A third point of attention is addressing optimality of these schedulers. Parameterizing the earliness function \eqref{eq:earliness} is not always trivial. Even so, finding a scheduler that minimizes the interventions is still an open problem. Priced TGA could be used, but their undecidability for games with three clocks has been proven by \cite{bouyer2006improved}, putting a roadblock in that direction. Approximate solutions using stochastic priced TGA \citep{david2015uppaal} are currently being explored.

\bibliography{mybib} 

\end{document}

%% file: tikzsettings.tex
\usepackage{pgfplots}
\usepackage{grffile}
\pgfplotsset{compat=newest}
\usetikzlibrary{arrows, arrows.meta, decorations.markings, patterns}
\usepgfplotslibrary{patchplots, groupplots}
\usetikzlibrary{shapes, positioning, fit, backgrounds}
\usepgfplotslibrary{fillbetween}

\usetikzlibrary{overlay-beamer-styles}
\usetikzlibrary{calc}

\usepackage{xcolor}
\usepackage{calc}

\definecolor{tudcyan}{RGB}{0,166,214}
\definecolor{tudmagenta}{RGB}{109,23,127}
\definecolor{tudpurple}{RGB}{29,28,115}
\definecolor{tudgraygreen}{RGB}{107,134,137}

\colorlet{lighttudcyan}{tudcyan!20}
\colorlet{lighttudmagenta}{tudmagenta!20}

\newlength{\hatchspread}
\newlength{\hatchthickness}
\newlength{\hatchshift}
\newcommand{\hatchcolor}{}
\tikzset{hatchspread/.code={\setlength{\hatchspread}{#1}},
	hatchthickness/.code={\setlength{\hatchthickness}{#1}},
	hatchshift/.code={\setlength{\hatchshift}{#1}},
	hatchcolor/.code={\renewcommand{\hatchcolor}{#1}}}
\tikzset{hatchspread=7pt,
	hatchthickness=0.5pt,
	hatchshift=0pt,
	hatchcolor=black}
\pgfdeclarepatternformonly[\hatchspread,\hatchthickness,\hatchshift,\hatchcolor]
{custom north west lines}
{\pgfqpoint{\dimexpr-2\hatchthickness}{\dimexpr-2\hatchthickness}}
{\pgfqpoint{\dimexpr\hatchspread+2\hatchthickness}{\dimexpr\hatchspread+2\hatchthickness}}
{\pgfqpoint{\dimexpr\hatchspread}{\dimexpr\hatchspread}}
{
	\pgfsetlinewidth{\hatchthickness}
	\pgfpathmoveto{\pgfqpoint{0pt}{\dimexpr\hatchspread+\hatchshift}}
	\pgfpathlineto{\pgfqpoint{\dimexpr\hatchspread+0.15pt+\hatchshift}{-0.15pt}}
	\ifdim \hatchshift > 0pt
	\pgfpathmoveto{\pgfqpoint{0pt}{\hatchshift}}
	\pgfpathlineto{\pgfqpoint{\dimexpr0.15pt+\hatchshift}{-0.15pt}}
	\fi
	\pgfsetstrokecolor{\hatchcolor}
	\pgfusepath{stroke}
}

\def\centerarc[#1](#2)(#3:#4:#5)
{ \draw[#1] ($(#2)+({#5*cos(#3)},{#5*sin(#3)})$) arc (#3:#4:#5); }

%% file: transition.tex
\begin{tikzpicture}

\definecolor{color0}{rgb}{0.12156862745098,0.466666666666667,0.705882352941177}
\definecolor{color1}{rgb}{1,0.498039215686275,0.0549019607843137}

\begin{axis}[
tick align=outside,
tick pos=left,
x grid style={white!69.01960784313725!black},
xmin=5, xmax=20,
xtick style={color=black},
xtick={6,8,10,12,14,16,18},
minor xtick={7, 9, 11, 13, 15, 17, 19},
xticklabel=$\Qs_{\pgfmathprintnumber{\tick}}$,
ytick={6,8,10,12,14,16,18},
minor ytick={7, 9, 11, 13, 15, 17, 19},
yticklabel=$\Qs_{\pgfmathprintnumber{\tick}}$,
y grid style={white!69.01960784313725!black},
ymin=5, ymax=20,
ytick style={color=black},
xlabel=from,
ylabel=to
]
\addplot [only marks, mark=x, mark size=5pt, draw=color0, colormap/viridis]
table{%
x                      y
6 6
6 7
6 8
6 9
6 10
6 11
6 12
7 6
7 7
7 8
7 9
7 10
8 8
8 9
8 6
8 7
9 6
9 7
9 8
9 9
9 10
9 11
9 12
9 13
9 14
9 15
9 16
9 17
9 18
9 19
10 6
10 7
10 8
10 9
10 10
10 11
10 12
10 13
10 14
10 15
10 16
10 17
10 18
10 19
11 6
11 7
11 8
11 9
11 10
11 11
11 12
11 13
11 14
11 15
11 16
11 17
11 18
11 19
12 6
12 7
12 8
12 9
12 10
12 11
12 12
12 13
12 14
12 15
12 16
12 17
12 18
12 19
13 6
13 7
13 8
13 9
13 10
13 11
13 12
13 13
13 14
13 15
13 16
13 17
13 18
13 19
14 6
14 7
14 8
14 9
14 10
14 11
14 12
14 13
14 14
14 15
14 16
14 17
14 18
14 19
15 6
15 7
15 8
15 9
15 10
15 11
15 12
15 13
15 14
15 15
15 16
15 17
15 18
15 19
16 6
16 7
16 8
16 9
16 10
16 11
16 12
16 13
16 14
16 15
16 16
16 17
16 18
16 19
17 6
17 7
17 8
17 9
17 10
17 11
17 12
17 13
17 14
17 15
17 16
17 17
17 18
17 19
18 6
18 7
18 8
18 9
18 10
18 11
18 12
18 13
18 14
18 15
18 16
18 17
18 18
19 6
19 7
19 8
19 9
19 10
19 11
19 12
19 13
19 14
19 15
19 16
19 17
19 18
19 19
};
\addplot [only marks, mark=o, mark size=5pt, draw=color1, fill=none, colormap/viridis]
table{%
x                      y
6 6
6 7
7 8
7 7
8 8
8 9
9 8
9 9
9 10
9 11
10 9
10 10
10 11
10 12
11 10
11 11
11 12
11 13
11 14
12 12
12 13
12 14
12 15
12 16
13 12
13 13
13 14
13 15
13 16
13 17
13 18
13 19
14 13
14 14
14 15
14 16
14 17
14 18
14 19
15 14
15 15
15 16
15 17
15 18
15 19
16 15
16 16
16 17
16 18
16 19
17 16
17 17
17 18
17 19
18 17
18 18
18 19
19 18
19 19
};
\end{axis}

\end{tikzpicture}

%% file: out1.tex
\begin{tikzpicture}

\definecolor{color0}{rgb}{0.12156862745098,0.466666666666667,0.705882352941177}
\definecolor{color1}{rgb}{1,0.498039215686275,0.0549019607843137}
\definecolor{color2}{rgb}{0.172549019607843,0.627450980392157,0.172549019607843}
\definecolor{color3}{rgb}{0.83921568627451,0.152941176470588,0.156862745098039}

\begin{groupplot}[group style={group size=1 by 2}]
\nextgroupplot[
height=4cm,
width=\linewidth,
tick align=outside,
tick pos=left,
x grid style={white!69.01960784313725!black},
xmin=0, xmax=1,
xtick style={color=black},
y grid style={white!69.01960784313725!black},
ymin=-1.20935840674518, ymax=1.70888495960873,
ytick style={color=black},
ylabel=State,
grid,
every axis plot/.append style={thick}
]
\addplot [thick, color0]
table {%
0 1
0.01 1.12496001035328
0.02 1.22175052348325
0.03 1.29246744007615
0.04 1.33903245906326
0.05 1.36320731110431
0.06 1.36660681657438
0.07 1.35071086561813
0.08 1.31687540935607
0.09 1.27404460738092
0.1 1.23027696724082
0.11 1.18551275008546
0.12 1.13969174603218
0.13 1.09275333229454
0.14 1.04463652054371
0.15 0.995279994588626
0.16 0.944622139359565
0.17 0.894738213445065
0.18 0.847582935216131
0.19 0.802929158780393
0.2 0.76056900631351
0.21 0.720312257066908
0.22 0.681984870851297
0.23 0.645427634799877
0.24 0.610494923146195
0.25 0.577282295635829
0.26 0.545844726908116
0.27 0.515996161509777
0.28 0.487564182762317
0.29 0.460388893701779
0.3 0.434321888738108
0.31 0.409225308579947
0.32 0.384970971585052
0.33 0.3614395752601
0.34 0.339049142568022
0.35 0.318205882642892
0.36 0.29878749321905
0.37 0.280682255434972
0.38 0.263788149376305
0.39 0.248012043638685
0.4 0.233268952742561
0.41 0.219461566645725
0.42 0.206487389931746
0.43 0.194264356522281
0.44 0.182716629146755
0.45 0.171774086120368
0.46 0.161371850033979
0.47 0.151449854900703
0.48 0.141952448589535
0.49 0.132971012456526
0.5 0.124595814362716
0.51 0.116776048933716
0.52 0.109465176484625
0.53 0.102620567145193
0.54 0.0962031746261584
0.55 0.0901772371615508
0.56 0.0845100033666405
0.57 0.0791714809390562
0.58 0.0741574679119865
0.59 0.0694602993915816
0.6 0.065048632692512
0.61 0.0608935093704127
0.62 0.0569681584889979
0.63 0.0532478159676517
0.64 0.0497095586831346
0.65 0.0463321521086512
0.66 0.0431499320492858
0.67 0.0401972637714388
0.68 0.0374554277702948
0.69 0.0349073084627114
0.7 0.0325372599776016
0.71 0.0303309831678872
0.72 0.0282754129091023
0.73 0.0263600635863336
0.74 0.0245739684088754
0.75 0.0229049708215132
0.76 0.0213418697764675
0.77 0.0198743405249915
0.78 0.0184928619277196
0.79 0.0171886497443959
0.8 0.0159535954082801
0.81 0.0147802098314887
0.82 0.0136799950785085
0.83 0.0126642723730713
0.84 0.0117261877987339
0.85 0.0108594760168065
0.86 0.0100584110133049
0.87 0.00931776096542355
0.88 0.00863226223790712
0.89 0.00799672172921792
0.9 0.00740665035144469
0.91 0.00685792170735264
0.92 0.0063467419264162
0.93 0.00586962199655969
0.94 0.00542335238462002
0.95 0.00500497975571387
0.96 0.00461178561743571
0.97 0.00424126672925057
0.98 0.0038911171306849
0.99 0.00356485848057198
};
\addplot [thick, color1, dashed]
table {%
0 -1
0.01 -0.628531614278419
0.02 -0.272819789811235
0.03 0.0680105368102824
0.04 0.394784505275102
0.05 0.708280380198291
0.06 1.00923240114699
0.07 1.29833344791917
0.08 1.57623753386537
0.09 1.39965612595167
0.1 1.23093380590185
0.11 1.06969566002093
0.12 0.915585404526165
0.13 0.768264480439955
0.14 0.627411192503331
0.15 0.492719890000291
0.16 0.363900187482423
0.17 0.394758964949638
0.18 0.424712048095821
0.19 0.453794148614565
0.2 0.482039074819436
0.21 0.509479708212628
0.22 0.536147986833455
0.23 0.562074894585117
0.24 0.587290455816503
0.25 0.51585363029868
0.26 0.447635315080927
0.27 0.382473790613377
0.28 0.320215918801629
0.29 0.260716677212218
0.3 0.203838719943006
0.31 0.149451963539933
0.32 0.0974331964466362
0.33 0.0476657105712461
0.34 0.0683696972037464
0.35 0.0883463431486386
0.36 0.10762549462566
0.37 0.126235961693792
0.38 0.144205531629237
0.39 0.161560984742757
0.4 0.178328112258206
0.41 0.155327544574723
0.42 0.133370335149148
0.43 0.112403162322638
0.44 0.0923756063532
0.45 0.073239986236559
0.46 0.0549512062921872
0.47 0.037466611891326
0.48 0.0207458537463376
0.49 0.0243171371781004
0.5 0.0277867664922253
0.51 0.031157592877683
0.52 0.0344325058905174
0.53 0.0376144166465191
0.54 0.04070624307953
0.55 0.0437108970608259
0.56 0.0466312731936369
0.57 0.0494702391138483
0.58 0.0410538857129209
0.59 0.0330127890946334
0.6 0.0253277205141344
0.61 0.0179805040974952
0.62 0.0109539572070797
0.63 0.00423183440783366
0.64 -0.00220122519749133
0.65 -0.00835974748019622
0.66 -0.0056204487305411
0.67 -0.00298251883544736
0.68 -0.000441833874165026
0.69 0.00200559363184876
0.7 0.00436361575077632
0.71 0.0066359505913424
0.72 0.00882618411304347
0.73 0.00691482587033945
0.74 0.00509055911151918
0.75 0.00334873342325216
0.76 0.00168496915476428
0.77 9.51408844768453e-05
0.78 -0.00142463802872056
0.79 -0.00287802988530295
0.8 -0.00426848614620376
0.81 -0.0055992600601693
0.82 -0.00458083526911666
0.83 -0.00360140601517921
0.84 -0.00265937417546856
0.85 -0.00175319591895746
0.86 -0.000881381122333517
0.87 -4.2492642105272e-05
0.88 -0.000338557429596601
0.89 -0.000620276683687849
0.9 -0.000888513871679865
0.91 -0.00114407433775491
0.92 -0.00138770940354669
0.93 -0.00162012016290487
0.94 -0.00184196099451112
0.95 -0.0020538428141248
0.96 -0.00225633608650885
0.97 -0.00244997361549898
0.98 -0.00263525312921925
0.99 -0.00230102476553499
};
\addplot [thick, color2, dotted]
table {%
0 1
0.01 0.606207875296529
0.02 0.257999404098999
0.03 -0.0479609039274114
0.04 -0.314753201111609
0.05 -0.54522243778968
0.06 -0.741996878602843
0.07 -0.907505131500304
0.08 -1.04399181045578
0.09 -1.02551685337716
0.1 -1.01410737670912
0.11 -1.00943626978134
0.12 -1.01119124602351
0.13 -1.01907422313822
0.14 -1.0328007246639
0.15 -1.05209930263549
0.16 -1.07671098100182
0.17 -1.00921599899927
0.18 -0.944322656297343
0.19 -0.881773682448377
0.2 -0.8213350755652
0.21 -0.762794076324705
0.22 -0.705957314805982
0.23 -0.650649115586347
0.24 -0.596709947739939
0.25 -0.558601921021691
0.26 -0.525864387710937
0.27 -0.498151403091086
0.28 -0.475140985434099
0.29 -0.456533362105378
0.3 -0.442049349674533
0.31 -0.431428857479749
0.32 -0.424429504941243
0.33 -0.420825343698071
0.34 -0.389634216963006
0.35 -0.359667282314424
0.36 -0.330793014562182
0.37 -0.30289221177278
0.38 -0.27585690282191
0.39 -0.249589349054576
0.4 -0.224001132072023
0.41 -0.208312397601078
0.42 -0.194688552608905
0.43 -0.182989702361124
0.44 -0.17308606720566
0.45 -0.164857217047559
0.46 -0.15819136572019
0.47 -0.152984720460891
0.48 -0.149140882088044
0.49 -0.138010479081738
0.5 -0.127549216918201
0.51 -0.117695931512949
0.52 -0.108394801243777
0.53 -0.0995948903808535
0.54 -0.091249731062152
0.55 -0.0833169405827134
0.56 -0.0757578710360906
0.57 -0.0685372885936459
0.58 -0.0634236189179321
0.59 -0.0590789825681847
0.6 -0.055450843424709
0.61 -0.0524904918188153
0.62 -0.050152753367645
0.63 -0.0483957206755102
0.64 -0.0471805060678875
0.65 -0.0464710136732712
0.66 -0.0427229668545567
0.67 -0.0391692275483032
0.68 -0.0357894728160006
0.69 -0.0325652664980685
0.7 -0.0294798927233253
0.71 -0.0265182037251796
0.72 -0.0236664807529558
0.73 -0.0217642108368373
0.74 -0.0201124948311713
0.75 -0.018692962829048
0.76 -0.0174886557568209
0.77 -0.0164839139134774
0.78 -0.0156642744841713
0.79 -0.0150163772966976
0.8 -0.0145278781497138
0.81 -0.0141873690965834
0.82 -0.012932202875667
0.83 -0.0117467941481553
0.84 -0.0106237609949764
0.85 -0.00955641036820612
0.86 -0.00853867714831124
0.87 -0.00756506844551197
0.88 -0.00689023597556017
0.89 -0.00629347243278422
0.9 -0.00576859897101524
0.91 -0.0053099359289826
0.92 -0.0049122620757173
0.93 -0.00457077720544644
0.94 -0.00428106780559407
0.95 -0.00403907554436436
0.96 -0.00384106834535063
0.97 -0.00368361383584449
0.98 -0.00356355497315755
0.99 -0.00319800030429685
};
\addplot [thick, color3, dash dot]
table {%
0 -1
0.01 -0.933263257095463
0.02 -0.857419449669665
0.03 -0.772735302498923
0.04 -0.67947770497061
0.05 -0.57791257522667
0.06 -0.468303873541117
0.07 -0.350912749773179
0.08 -0.225996811184751
0.09 -0.192100400640498
0.1 -0.166030650570817
0.11 -0.147392478432694
0.12 -0.135810262402798
0.13 -0.130926860883292
0.14 -0.132402681631979
0.15 -0.139914797954618
0.16 -0.15315610953291
0.17 -0.136760256321934
0.18 -0.118563979349705
0.19 -0.0986739116620177
0.2 -0.0771898592055071
0.21 -0.0542052684500258
0.22 -0.0298076600961183
0.23 -0.00407903144147262
0.24 0.0229037702215291
0.25 0.0295330217179168
0.26 0.0335274039303982
0.27 0.0350061638423509
0.28 0.0340837619729747
0.29 0.0308700245048176
0.3 0.0254702950134991
0.31 0.0179855852324131
0.32 0.00851272436193026
0.33 -0.00285549349847056
0.34 -0.000559102238308198
0.35 0.00294532235127451
0.36 0.00758783853199035
0.37 0.0133028097067354
0.38 0.0200286196817686
0.39 0.0277074079553585
0.4 0.0362848235479156
0.41 0.036889858289936
0.42 0.0367202903602988
0.43 0.0358091009795053
0.44 0.0341881455727524
0.45 0.031888169997091
0.46 0.0289388292754065
0.47 0.0253687084400293
0.48 0.021205345131096
0.49 0.0209036627915286
0.5 0.0208963691443774
0.51 0.0211648939993189
0.52 0.021691928836292
0.53 0.0224613358247452
0.54 0.0234580637160738
0.55 0.0246680700721645
0.56 0.0260782493359033
0.57 0.0276763662889471
0.58 0.0269422633977657
0.59 0.0259360678495694
0.6 0.0246691775382994
0.61 0.0231526205721986
0.62 0.0213970584141591
0.63 0.019412790137707
0.64 0.0172097576385064
0.65 0.0147975516578909
0.66 0.0141384048337186
0.67 0.0136536787442746
0.68 0.0133332035682299
0.69 0.0131674415067358
0.7 0.0131474446570165
0.71 0.0132648158705607
0.72 0.0135116723733672
0.73 0.0129762442745603
0.74 0.0123957196140753
0.75 0.0117714729981974
0.76 0.0111048886331411
0.77 0.0103973536580188
0.78 0.00965025232536286
0.79 0.00886496094443635
0.8 0.00804284351031737
0.81 0.00718524794881159
0.82 0.00681270967024818
0.83 0.00650586222967489
0.84 0.00626088459674104
0.85 0.00607419204275266
0.86 0.00594242046222231
0.87 0.00586241180070657
0.88 0.00558297550117112
0.89 0.00530527303828117
0.9 0.00502888327249851
0.91 0.004753435264853
0.92 0.00447860340788442
0.93 0.00420410299638491
0.94 0.00392968619969807
0.95 0.0036551384005916
0.96 0.00338027486870558
0.97 0.00310493773931309
0.98 0.00282899327063236
0.99 0.00266850141477946
};

\nextgroupplot[
height=4cm,
width=\linewidth,
tick align=outside,
tick pos=left,
x grid style={white!69.01960784313725!black},
xmin=0, xmax=1,
xtick style={color=black},
y grid style={white!69.01960784313725!black},
ymin=-2.44463105992804, ymax=11.1090360654144,
ytick style={color=black},
ylabel=Control input,
xlabel=$t$,
grid,
every axis plot/.append style={thick}
]
\addplot [thick, color0, const plot mark left]
table {%
0 6.15093290959004
0.01 6.15093290959004
0.02 6.15093290959004
0.03 6.15093290959004
0.04 6.15093290959004
0.05 6.15093290959004
0.06 6.15093290959004
0.07 6.15093290959004
0.08 -1.82855528150339
0.09 -1.82855528150339
0.1 -1.82855528150339
0.11 -1.82855528150339
0.12 -1.82855528150339
0.13 -1.82855528150339
0.14 -1.82855528150339
0.15 -1.82855528150339
0.16 0.94130641846495
0.17 0.94130641846495
0.18 0.94130641846495
0.19 0.94130641846495
0.2 0.94130641846495
0.21 0.94130641846495
0.22 0.94130641846495
0.23 0.94130641846495
0.24 -0.783854460249914
0.25 -0.783854460249914
0.26 -0.783854460249914
0.27 -0.783854460249914
0.28 -0.783854460249914
0.29 -0.783854460249914
0.3 -0.783854460249914
0.31 -0.783854460249914
0.32 -0.783854460249914
0.33 0.444479036124188
0.34 0.444479036124188
0.35 0.444479036124188
0.36 0.444479036124188
0.37 0.444479036124188
0.38 0.444479036124188
0.39 0.444479036124188
0.4 -0.260259455314928
0.41 -0.260259455314928
0.42 -0.260259455314928
0.43 -0.260259455314928
0.44 -0.260259455314928
0.45 -0.260259455314928
0.46 -0.260259455314928
0.47 -0.260259455314928
0.48 0.0914712507329042
0.49 0.0914712507329042
0.5 0.0914712507329042
0.51 0.0914712507329042
0.52 0.0914712507329042
0.53 0.0914712507329042
0.54 0.0914712507329042
0.55 0.0914712507329042
0.56 0.0914712507329042
0.57 -0.109442209289144
0.58 -0.109442209289144
0.59 -0.109442209289144
0.6 -0.109442209289144
0.61 -0.109442209289144
0.62 -0.109442209289144
0.63 -0.109442209289144
0.64 -0.109442209289144
0.65 0.0458190740945032
0.66 0.0458190740945032
0.67 0.0458190740945032
0.68 0.0458190740945032
0.69 0.0458190740945032
0.7 0.0458190740945032
0.71 0.0458190740945032
0.72 -0.0264977223731038
0.73 -0.0264977223731038
0.74 -0.0264977223731038
0.75 -0.0264977223731038
0.76 -0.0264977223731038
0.77 -0.0264977223731038
0.78 -0.0264977223731038
0.79 -0.0264977223731038
0.8 -0.0264977223731038
0.81 0.0147144438900374
0.82 0.0147144438900374
0.83 0.0147144438900374
0.84 0.0147144438900374
0.85 0.0147144438900374
0.86 0.0147144438900374
0.87 -0.00512059878561054
0.88 -0.00512059878561054
0.89 -0.00512059878561054
0.9 -0.00512059878561054
0.91 -0.00512059878561054
0.92 -0.00512059878561054
0.93 -0.00512059878561054
0.94 -0.00512059878561054
0.95 -0.00512059878561054
0.96 -0.00512059878561054
0.97 -0.00512059878561054
0.98 0.00407638581721532
0.99 0.00407638581721532
};
\addplot [thick, color1, dashed, const plot mark left]
table {%
0 10.4929602869898
0.01 10.4929602869898
0.02 10.4929602869898
0.03 10.4929602869898
0.04 10.4929602869898
0.05 10.4929602869898
0.06 10.4929602869898
0.07 10.4929602869898
0.08 3.00889884476741
0.09 3.00889884476741
0.1 3.00889884476741
0.11 3.00889884476741
0.12 3.00889884476741
0.13 3.00889884476741
0.14 3.00889884476741
0.15 3.00889884476741
0.16 0.955272134754585
0.17 0.955272134754585
0.18 0.955272134754585
0.19 0.955272134754585
0.2 0.955272134754585
0.21 0.955272134754585
0.22 0.955272134754585
0.23 0.955272134754585
0.24 0.726133521167948
0.25 0.726133521167948
0.26 0.726133521167948
0.27 0.726133521167948
0.28 0.726133521167948
0.29 0.726133521167948
0.3 0.726133521167948
0.31 0.726133521167948
0.32 0.726133521167948
0.33 0.220141808713698
0.34 0.220141808713698
0.35 0.220141808713698
0.36 0.220141808713698
0.37 0.220141808713698
0.38 0.220141808713698
0.39 0.220141808713698
0.4 0.236019997506444
0.41 0.236019997506444
0.42 0.236019997506444
0.43 0.236019997506444
0.44 0.236019997506444
0.45 0.236019997506444
0.46 0.236019997506444
0.47 0.236019997506444
0.48 0.0993924451458139
0.49 0.0993924451458139
0.5 0.0993924451458139
0.51 0.0993924451458139
0.52 0.0993924451458139
0.53 0.0993924451458139
0.54 0.0993924451458139
0.55 0.0993924451458139
0.56 0.0993924451458139
0.57 0.075972080194554
0.58 0.075972080194554
0.59 0.075972080194554
0.6 0.075972080194554
0.61 0.075972080194554
0.62 0.075972080194554
0.63 0.075972080194554
0.64 0.075972080194554
0.65 0.024455665797971
0.66 0.024455665797971
0.67 0.024455665797971
0.68 0.024455665797971
0.69 0.024455665797971
0.7 0.024455665797971
0.71 0.024455665797971
0.72 0.0227206316893958
0.73 0.0227206316893958
0.74 0.0227206316893958
0.75 0.0227206316893958
0.76 0.0227206316893958
0.77 0.0227206316893958
0.78 0.0227206316893958
0.79 0.0227206316893958
0.8 0.0227206316893958
0.81 0.00509751373260344
0.82 0.00509751373260344
0.83 0.00509751373260344
0.84 0.00509751373260344
0.85 0.00509751373260344
0.86 0.00509751373260344
0.87 0.0054744526445723
0.88 0.0054744526445723
0.89 0.0054744526445723
0.9 0.0054744526445723
0.91 0.0054744526445723
0.92 0.0054744526445723
0.93 0.0054744526445723
0.94 0.0054744526445723
0.95 0.0054744526445723
0.96 0.0054744526445723
0.97 0.0054744526445723
0.98 5.69978976194107e-05
0.99 5.69978976194107e-05
};
\end{groupplot}

\end{tikzpicture}

%% file: out2.tex
\begin{tikzpicture}

\definecolor{color0}{rgb}{0.12156862745098,0.466666666666667,0.705882352941177}
\definecolor{color1}{rgb}{1,0.498039215686275,0.0549019607843137}
\definecolor{color2}{rgb}{0.172549019607843,0.627450980392157,0.172549019607843}
\definecolor{color3}{rgb}{0.83921568627451,0.152941176470588,0.156862745098039}

\begin{groupplot}[group style={group size=1 by 2}]
\nextgroupplot[
height=4cm,
width=\linewidth,
tick align=outside,
tick pos=left,
x grid style={white!69.01960784313725!black},
xmin=0, xmax=1,
xtick style={color=black},
y grid style={white!69.01960784313725!black},
ymin=-4.09208910843289, ymax=4.42843657225479,
ytick style={color=black},
ylabel=State,
grid
]
\addplot [thick, color0]
table {%
0 1
0.01 0.987182288245613
0.02 0.968992196038882
0.03 0.934108543459323
0.04 0.882747631385991
0.05 0.81506304518069
0.06 0.740139371927793
0.07 0.666849837734002
0.08 0.595003705943381
0.09 0.52443064123142
0.1 0.454978832835724
0.11 0.386513283461138
0.12 0.318914249636363
0.13 0.254991121607436
0.14 0.197209176413938
0.15 0.14498053927948
0.16 0.0977639475229923
0.17 0.0550608643558365
0.18 0.0164119135961952
0.19 -0.0186063913064817
0.2 -0.0496619476907561
0.21 -0.076355263746132
0.22 -0.0989611360799605
0.23 -0.117726393825803
0.24 -0.132872280445505
0.25 -0.144596630943544
0.26 -0.1539322705206
0.27 -0.161899976157158
0.28 -0.168654000426048
0.29 -0.17433718969739
0.3 -0.179081932511678
0.31 -0.182773901730296
0.32 -0.18526706163908
0.33 -0.186632707470746
0.34 -0.18693441780703
0.35 -0.186228718566785
0.36 -0.184565689408236
0.37 -0.181989517420895
0.38 -0.178886325568651
0.39 -0.175631998387806
0.4 -0.172249913593714
0.41 -0.16876178363095
0.42 -0.165187797241822
0.43 -0.161546749459302
0.44 -0.157856160983461
0.45 -0.154035872246032
0.46 -0.149991508795834
0.47 -0.145717702119924
0.48 -0.141208251388051
0.49 -0.136456205175909
0.5 -0.131453934767857
0.51 -0.126425252521505
0.52 -0.121592082629797
0.53 -0.116944360654047
0.54 -0.112473097338704
0.55 -0.108170288879307
0.56 -0.103997860149136
0.57 -0.0999117700736359
0.58 -0.0958964302646539
0.59 -0.0919370966424653
0.6 -0.0880198038512914
0.61 -0.0841313045414644
0.62 -0.0802590131355862
0.63 -0.0763909537265939
0.64 -0.0726412494355069
0.65 -0.0691205015040594
0.66 -0.065813029029798
0.67 -0.0627045818757168
0.68 -0.05978222162493
0.69 -0.057009073301212
0.7 -0.0543468758392683
0.71 -0.0517822465204538
0.72 -0.0493026972763038
0.73 -0.0468965624605655
0.74 -0.0445529323428797
0.75 -0.0422615918591798
0.76 -0.0400129641919009
0.77 -0.0377980587879953
0.78 -0.0356785972662735
0.79 -0.0337146338953892
0.8 -0.0318958650203916
0.81 -0.0302129538929173
0.82 -0.0286574498638414
0.83 -0.0272003021550384
0.84 -0.0258120864861378
0.85 -0.0244848503952794
0.86 -0.0232111885427007
0.87 -0.0219841983018906
0.88 -0.0207974388955845
0.89 -0.0196448937874778
0.9 -0.0185452419400595
0.91 -0.0175172189211661
0.92 -0.0165555678005617
0.93 -0.0156555013200671
0.94 -0.0148126626633096
0.95 -0.0140230895221116
0.96 -0.0132774931385494
0.97 -0.0125661047055185
0.98 -0.0118847452194397
0.99 -0.0112295192749018
};
\addplot [thick, color1, dashed]
table {%
0 2
0.01 1.93693382368946
0.02 0.437849648310537
0.03 -1.00023887552345
0.04 -2.38020778873797
0.05 -3.70479248658345
0.06 -3.33082556062054
0.07 -2.97307302510739
0.08 -2.63074913324232
0.09 -2.30310583954922
0.1 -1.98943103167037
0.11 -1.68904684189749
0.12 -1.40130803512933
0.13 -1.46902881874773
0.14 -1.53435610113372
0.15 -1.59742592359188
0.16 -1.6583655689894
0.17 -1.71729415109772
0.18 -1.77432316154553
0.19 -1.82955697757428
0.2 -1.60697148759267
0.21 -1.39393437914721
0.22 -1.18999878484646
0.23 -0.994738780928798
0.24 -0.80774846491785
0.25 -0.628641069996598
0.26 -0.679057263353589
0.27 -0.727666144776441
0.28 -0.77455945578607
0.29 -0.819823885863496
0.3 -0.863541358230811
0.31 -0.784435718686694
0.32 -0.708835270603494
0.33 -0.636574481196076
0.34 -0.567495679295547
0.35 -0.501448699318157
0.36 -0.438290540299749
0.37 -0.377885039462292
0.38 -0.386063262195003
0.39 -0.394075012545015
0.4 -0.401928679932839
0.41 -0.409632154206632
0.42 -0.417192855683024
0.43 -0.424617763222944
0.44 -0.43191344047879
0.45 -0.378707768104239
0.46 -0.327823373219252
0.47 -0.279148771564927
0.48 -0.232577963587377
0.49 -0.188010170170847
0.5 -0.145349580992579
0.51 -0.154939972140575
0.52 -0.164221656431962
0.53 -0.173208834543149
0.54 -0.181915043730174
0.55 -0.19035318633101
0.56 -0.171759228556887
0.57 -0.154002539869989
0.58 -0.137042323885638
0.59 -0.120839838591786
0.6 -0.105358293306273
0.61 -0.0905627508961972
0.62 -0.0764200349856266
0.63 -0.0628986418925621
0.64 -0.0677203325378468
0.65 -0.0723872393096069
0.66 -0.0769060122062542
0.67 -0.0812830285538879
0.68 -0.0855244012460442
0.69 -0.0766013874861369
0.7 -0.0680806679567738
0.71 -0.0599423994135013
0.72 -0.0521677581677827
0.73 -0.044738887282328
0.74 -0.0376388465980855
0.75 -0.0308515654347049
0.76 -0.0243617978156334
0.77 -0.018155080077759
0.78 -0.0224163046589668
0.79 -0.0265214771874894
0.8 -0.0304775791763959
0.81 -0.0342912850199078
0.82 -0.0379689735489972
0.83 -0.0332950951088732
0.84 -0.0288292385465126
0.85 -0.0245611032779788
0.86 -0.020480921529673
0.87 -0.0165794304651483
0.88 -0.0128478458286205
0.89 -0.00927783701876507
0.9 -0.0106805217552608
0.91 -0.0120346171067827
0.92 -0.013342220480525
0.93 -0.0146053446286261
0.94 -0.0158259201220866
0.95 -0.01700579784096
0.96 -0.014483584203935
0.97 -0.0120720902402391
0.98 -0.00976584638810332
0.99 -0.00755966596352724
};
\addplot [thick, color2, dotted]
table {%
0 3
0.01 3.12772045601993
0.02 2.59227827735137
0.03 2.01216008236704
0.04 1.38900330230396
0.05 0.724406427811791
0.06 0.650472337429419
0.07 0.587612735477621
0.08 0.535456911901085
0.09 0.493638402344704
0.1 0.46179594447187
0.11 0.439574290375093
0.12 0.426624890206504
0.13 0.432057731251968
0.14 0.426896483295578
0.15 0.411959535492575
0.16 0.387999933870158
0.17 0.355710676175397
0.18 0.315729573546173
0.19 0.268643714649678
0.2 0.297666459900957
0.21 0.330372218418773
0.22 0.366785342857321
0.23 0.406912384723214
0.24 0.450744352375634
0.25 0.498258743153546
0.26 0.474845164696657
0.27 0.44770661212947
0.28 0.41709549346182
0.29 0.383245816931445
0.3 0.34637459164174
0.31 0.34064782305602
0.32 0.33672161304842
0.33 0.334565893699142
0.34 0.334147355310164
0.35 0.33542999917107
0.36 0.338375630466252
0.37 0.342944296981633
0.38 0.324203499310379
0.39 0.304996618694531
0.4 0.285356562872645
0.41 0.265313717752543
0.42 0.244896146184694
0.43 0.224129770752727
0.44 0.203038541882263
0.45 0.19791152631237
0.46 0.194748116181489
0.47 0.193462611228847
0.48 0.19397273581262
0.49 0.196199537130262
0.5 0.200067282899394
0.51 0.187457708177903
0.52 0.174751368029123
0.53 0.161944975175006
0.54 0.149036024830333
0.55 0.136022704475226
0.56 0.129769643375559
0.57 0.124428244069964
0.58 0.119950045722575
0.59 0.116289306552835
0.6 0.11340284261001
0.61 0.111249876755154
0.62 0.109791897159877
0.63 0.108992524680884
0.64 0.101174682538759
0.65 0.0934682744386093
0.66 0.0858579688138567
0.67 0.0783299829596556
0.68 0.0708719407673792
0.69 0.0671111580948416
0.7 0.0638761507483534
0.71 0.0611361259286355
0.72 0.0588622323562151
0.73 0.0570274295086598
0.74 0.0556063662151245
0.75 0.0545752679053247
0.76 0.0539118318647568
0.77 0.053595129898329
0.78 0.0492720517313455
0.79 0.044978054838548
0.8 0.0407049530076976
0.81 0.0364455077524586
0.82 0.0321933363575956
0.83 0.0304013146141446
0.84 0.0288965400538121
0.85 0.0276617898675153
0.86 0.0266809552076269
0.87 0.0259389643373473
0.88 0.0254217113902416
0.89 0.0251159903121894
0.9 0.0232223834745065
0.91 0.0213678582234306
0.92 0.0195472771536693
0.93 0.0177560170830007
0.94 0.0159899220213741
0.95 0.0142452602580129
0.96 0.0135006714792998
0.97 0.0129077401311303
0.98 0.0124574124638004
0.99 0.0121412183238843
};
\addplot [thick, color3, dash dot]
table {%
0 4
0.01 4.04113995040535
0.02 3.75460329641379
0.03 3.40452192691636
0.04 2.99414580178605
0.05 2.52656020532767
0.06 2.37657280599179
0.07 2.24423051484252
0.08 2.12864259095169
0.09 2.02896427165414
0.1 1.94439433112903
0.11 1.87417277375533
0.12 1.81757865439962
0.13 1.69727980364144
0.14 1.57664513429016
0.15 1.45564744110704
0.16 1.3342646872777
0.17 1.21247944310152
0.18 1.09027837802813
0.19 0.967651801284244
0.2 0.906829606966775
0.21 0.856877797163146
0.22 0.817226999268474
0.23 0.787338221190884
0.24 0.766701144405349
0.25 0.754832518159985
0.26 0.701180228700833
0.27 0.646211497805732
0.28 0.589981002037389
0.29 0.532541668249808
0.3 0.47394468763779
0.31 0.441575960520276
0.32 0.413085531643482
0.33 0.388272266571928
0.34 0.36694559592611
0.35 0.34892493478384
0.36 0.334039135541436
0.37 0.322125972181458
0.38 0.298129725369909
0.39 0.274040148316466
0.4 0.249860137811013
0.41 0.225592595892279
0.42 0.201240418804059
0.43 0.176806487422433
0.44 0.152293659005178
0.45 0.14130146279436
0.46 0.132685125448973
0.47 0.126325042243966
0.48 0.122107582058293
0.49 0.119924781952192
0.5 0.119674057614115
0.51 0.10987028636711
0.52 0.0997058388609555
0.53 0.0891995747751992
0.54 0.07836934100491
0.55 0.0672320294886412
0.56 0.0618310914061591
0.57 0.057235904437056
0.58 0.0534069150798669
0.59 0.0503064558914383
0.6 0.0478986562211468
0.61 0.0461493570042323
0.62 0.0450260294418684
0.63 0.0444976974017719
0.64 0.0405180036730833
0.65 0.036319187261058
0.66 0.0319134850438356
0.67 0.0273124175964968
0.68 0.0225268344295168
0.69 0.0205030984264949
0.7 0.0188450391903967
0.71 0.017535187850213
0.72 0.0165568664723245
0.73 0.0158941541250864
0.74 0.0155318541804096
0.75 0.0154554628272415
0.76 0.0156511387701004
0.77 0.0161056740844042
0.78 0.0144991055379774
0.79 0.0126928623839251
0.8 0.0106978105383356
0.81 0.00852420333436842
0.82 0.0061817185488159
0.83 0.00553258569393798
0.84 0.0050689408155361
0.85 0.00478201704680024
0.86 0.00466343646526748
0.87 0.00470519409878887
0.88 0.00489964244739997
0.89 0.00523947651812684
0.9 0.0046292130926719
0.91 0.00394898945714736
0.92 0.00320272370938681
0.93 0.00239410308261587
0.94 0.0015265985811106
0.95 0.000603478626817758
0.96 0.000452704069385982
0.97 0.000400820272743382
0.98 0.000443142948609335
0.99 0.000575196086798542
};

\nextgroupplot[
height=4cm,
width=\linewidth,
tick align=outside,
tick pos=left,
x grid style={white!69.01960784313725!black},
xmin=0, xmax=1,
xtick style={color=black},
y grid style={white!69.01960784313725!black},
ymin=-27.7046261826648, ymax=12.7084308424227,
ytick style={color=black},
grid,
xlabel=$t$,
ylabel=Control input
]
\addplot [thick, color0, const plot mark left]
table {%
0 -0
0.01 -25.8676690451608
0.02 -25.8676690451608
0.03 -25.8676690451608
0.04 -25.8676690451608
0.05 3.71705347415226
0.06 3.71705347415226
0.07 3.71705347415226
0.08 3.71705347415226
0.09 3.71705347415226
0.1 3.71705347415226
0.11 3.71705347415226
0.12 -2.45636688182623
0.13 -2.45636688182623
0.14 -2.45636688182623
0.15 -2.45636688182623
0.16 -2.45636688182623
0.17 -2.45636688182623
0.18 -2.45636688182623
0.19 2.50724275535033
0.2 2.50724275535033
0.21 2.50724275535033
0.22 2.50724275535033
0.23 2.50724275535033
0.24 2.50724275535033
0.25 -1.48363586686971
0.26 -1.48363586686971
0.27 -1.48363586686971
0.28 -1.48363586686971
0.29 -1.48363586686971
0.3 0.69783298354268
0.31 0.69783298354268
0.32 0.69783298354268
0.33 0.69783298354268
0.34 0.69783298354268
0.35 0.69783298354268
0.36 0.69783298354268
0.37 -0.487891605023439
0.38 -0.487891605023439
0.39 -0.487891605023439
0.4 -0.487891605023439
0.41 -0.487891605023439
0.42 -0.487891605023439
0.43 -0.487891605023439
0.44 0.597476416157487
0.45 0.597476416157487
0.46 0.597476416157487
0.47 0.597476416157487
0.48 0.597476416157487
0.49 0.597476416157487
0.5 -0.309151155993289
0.51 -0.309151155993289
0.52 -0.309151155993289
0.53 -0.309151155993289
0.54 -0.309151155993289
0.55 0.172183288664255
0.56 0.172183288664255
0.57 0.172183288664255
0.58 0.172183288664255
0.59 0.172183288664255
0.6 0.172183288664255
0.61 0.172183288664255
0.62 0.172183288664255
0.63 -0.146925720936619
0.64 -0.146925720936619
0.65 -0.146925720936619
0.66 -0.146925720936619
0.67 -0.146925720936619
0.68 0.0873860465439023
0.69 0.0873860465439023
0.7 0.0873860465439023
0.71 0.0873860465439023
0.72 0.0873860465439023
0.73 0.0873860465439023
0.74 0.0873860465439023
0.75 0.0873860465439023
0.76 0.0873860465439023
0.77 -0.0959469954507985
0.78 -0.0959469954507985
0.79 -0.0959469954507985
0.8 -0.0959469954507985
0.81 -0.0959469954507985
0.82 0.0518465680827264
0.83 0.0518465680827264
0.84 0.0518465680827264
0.85 0.0518465680827264
0.86 0.0518465680827264
0.87 0.0518465680827264
0.88 0.0518465680827264
0.89 -0.0347811714395533
0.9 -0.0347811714395533
0.91 -0.0347811714395533
0.92 -0.0347811714395533
0.93 -0.0347811714395533
0.94 -0.0347811714395533
0.95 0.0310683943904125
0.96 0.0310683943904125
0.97 0.0310683943904125
0.98 0.0310683943904125
0.99 0.0310683943904125
};
\addplot [thick, color1, const plot mark left, dashed]
table {%
0 0
0.01 10.8714737049187
0.02 10.8714737049187
0.03 10.8714737049187
0.04 10.8714737049187
0.05 2.31724744991551
0.06 2.31724744991551
0.07 2.31724744991551
0.08 2.31724744991551
0.09 2.31724744991551
0.1 2.31724744991551
0.11 2.31724744991551
0.12 -0.530110289691661
0.13 -0.530110289691661
0.14 -0.530110289691661
0.15 -0.530110289691661
0.16 -0.530110289691661
0.17 -0.530110289691661
0.18 -0.530110289691661
0.19 -1.20595528489004
0.2 -1.20595528489004
0.21 -1.20595528489004
0.22 -1.20595528489004
0.23 -1.20595528489004
0.24 -1.20595528489004
0.25 -0.396331319415761
0.26 -0.396331319415761
0.27 -0.396331319415761
0.28 -0.396331319415761
0.29 -0.396331319415761
0.3 -0.615531151035292
0.31 -0.615531151035292
0.32 -0.615531151035292
0.33 -0.615531151035292
0.34 -0.615531151035292
0.35 -0.615531151035292
0.36 -0.615531151035292
0.37 -0.285181838237504
0.38 -0.285181838237504
0.39 -0.285181838237504
0.4 -0.285181838237504
0.41 -0.285181838237504
0.42 -0.285181838237504
0.43 -0.285181838237504
0.44 -0.373487359474875
0.45 -0.373487359474875
0.46 -0.373487359474875
0.47 -0.373487359474875
0.48 -0.373487359474875
0.49 -0.373487359474875
0.5 -0.153303877430933
0.51 -0.153303877430933
0.52 -0.153303877430933
0.53 -0.153303877430933
0.54 -0.153303877430933
0.55 -0.181029062459216
0.56 -0.181029062459216
0.57 -0.181029062459216
0.58 -0.181029062459216
0.59 -0.181029062459216
0.6 -0.181029062459216
0.61 -0.181029062459216
0.62 -0.181029062459216
0.63 -0.0611205439884289
0.64 -0.0611205439884289
0.65 -0.0611205439884289
0.66 -0.0611205439884289
0.67 -0.0611205439884289
0.68 -0.0843449860585068
0.69 -0.0843449860585068
0.7 -0.0843449860585068
0.71 -0.0843449860585068
0.72 -0.0843449860585068
0.73 -0.0843449860585068
0.74 -0.0843449860585068
0.75 -0.0843449860585068
0.76 -0.0843449860585068
0.77 -0.017340684588268
0.78 -0.017340684588268
0.79 -0.017340684588268
0.8 -0.017340684588268
0.81 -0.017340684588268
0.82 -0.0373613162346576
0.83 -0.0373613162346576
0.84 -0.0373613162346576
0.85 -0.0373613162346576
0.86 -0.0373613162346576
0.87 -0.0373613162346576
0.88 -0.0373613162346576
0.89 -0.0142601379670003
0.9 -0.0142601379670003
0.91 -0.0142601379670003
0.92 -0.0142601379670003
0.93 -0.0142601379670003
0.94 -0.0142601379670003
0.95 -0.0194556233435994
0.96 -0.0194556233435994
0.97 -0.0194556233435994
0.98 -0.0194556233435994
0.99 -0.0194556233435994
};
\end{groupplot}

\end{tikzpicture}

%% file: trigger.tex
\begin{tikzpicture}

\definecolor{color0}{rgb}{0.12156862745098,0.466666666666667,0.705882352941177}
\definecolor{color1}{rgb}{1,0.498039215686275,0.0549019607843137}
\definecolor{color2}{rgb}{0.172549019607843,0.627450980392157,0.172549019607843}
\definecolor{color3}{rgb}{0.83921568627451,0.152941176470588,0.156862745098039}

\begin{axis}[
height=2.3cm,
width=\linewidth,
tick align=outside,
tick pos=left,
x grid style={white!69.01960784313725!black},
xmin=0, xmax=1,
xtick style={color=black},
xtick={0,0.2,0.4,0.6,0.8},
xticklabels={0.0,0.2,0.4,0.6,0.8},
y grid style={white!69.01960784313725!black},
ymin=0.5, ymax=2.5,
ytick style={color=black},
ytick={1, 2},
yticklabels={Plant 1, Plant 2},
axis x line=center,
axis y line=center,
every axis plot/.append style={thick}
]
\addplot [only marks, draw=color0, fill=none, mark=o, mark size=4pt, colormap/viridis]
table{%
x                      y
0.08 1
0.16 1
0.24 1
0.48 1
0.81 1
0.87 1
0.98 1
};
\addplot [only marks, fill=none, draw=color2, mark=o, mark size=4pt, colormap/viridis]
table{%
x                      y
0.3 2
0.37 2
0.55 2
0.63 2
0.68 2
};
\addplot [only marks, draw=color0, fill=none, mark=x, mark size=4pt, colormap/viridis]
table{%
x                      y
0 1
0.33 1
0.4 1
0.57 1
0.65 1
0.72 1
};
\addplot [only marks, draw=color2, fill=none, mark=x, mark size=4pt, colormap/viridis]
table{%
x                      y
0.01 2
0.05 2
0.12 2
0.19 2
0.25 2
0.44 2
0.5 2
0.77 2
0.82 2
0.89 2
0.95 2
};
\end{axis}

\end{tikzpicture}